\documentclass[article,lineno]{article}
\usepackage{filecontents}
\usepackage{bm}
\usepackage{natbib}
\usepackage{amsmath}
\usepackage{amssymb}
\usepackage{epsfig}
\usepackage{graphics}
\usepackage{graphicx,color}

\begin{document}

\title{A Bayesian GED-Gamma stochastic volatility model for return data: a marginal likelihood approach}

 \author{T. R. Santos \thanks{T. R. Santos
    Universidade Federal de Minas Gerais, Brazil.
    E-mail: thiagords@est.ufmg.br. The author received support from the FAPEMIG Foundation and CNPq-Brazil}\hspace{.2cm}\\
    Department of Statistics, Universidade Federal de Minas Gerais\\ \date{}
    }

 \date{}

\maketitle









\begin{abstract}
Several studies explore inferences based on stochastic volatility (SV) models, taking into
account the stylized facts of return data. The common problem is that the latent parameters of many volatility models are high-dimensional and analytically intractable, which
means inferences require approximations using, for example, the Markov Chain Monte Carlo or Laplace methods. Some SV models are expressed as a linear Gaussian state-space model that leads to a marginal likelihood, reducing the dimensionality of the problem. Others are not linearized, and the latent parameters are integrated out. However, these present a quite restrictive evolution equation. Thus, we propose a Bayesian GED-Gamma SV model with a direct marginal likelihood that is a product of the generalized Student's t-distributions in which the latent states are related across time through a stationary Gaussian evolution equation. Then, an approximation is made for the prior distribution of log-precision/volatility, without the need for model linearization. This also allows for the computation of the marginal likelihood function, where the high-dimensional latent states are integrated out and easily sampled in blocks using a smoothing procedure. In addition, extensions of our GED-Gamma model are easily made to incorporate skew heavy-tailed distributions. We use the Bayesian estimator for the inference of static parameters, and perform a simulation study on several properties of the estimator. Our results show that the proposed model can be reasonably estimated. Furthermore, we provide case studies of a Brazilian asset and the pound/dollar exchange rate to show the performance of our approach in terms of fit and prediction. \\

{\it Keywords: SV model, New sequential and smoothing procedures, Generalized Student's t-distribution, Non-Gaussian errors, Heavy tails, Skewness}
\end{abstract}

\section {Introduction}

There is evidence of non-Gaussianity, skewness, and heavy tails in the distribution of return data. Therefore, we need to choose
more flexible models in order to incorporate these stylized facts. Volatility is an important statistical measure,
representing the conditional variance of an underlying asset return,
 and plays a key role in finance \citep{bib:tsay2010}. It is also an important component of risk management,
 portfolio optimization, and options trading.

Since volatility is a latent component, its estimation calls for specific techniques and
suitable statistical inferences. Several models have been proposed for estimating the volatility of asset return data. For
example, \citet{bib:engle1982}
introduced an autoregressive conditional heteroskedasticity (ARCH) model, where volatility is a
function of past time series values. The generalized ARCH (GARCH) model was proposed by \citet{bib:bollereslev1986}, where
volatility can depend on its own past. \citet{bib:taylor1982} proposed
 a stochastic volatility (SV) model with an error term in its volatility equation, capturing
  some of the characteristics found in financial return time series in a better way than the GARCH model.

The usual approach to using a SV model is to use linearization to convert it into
a linear state-space model by a transformation in the return data. \citet{bib:harvey1994}, who adopted
this approach, presented
the quasi-maximum likelihood (QML) estimator from the classical perspective, considering
the innovation distribution to be approximately Gaussian. \citet{bib:danielsson} proposed the simulated
maximum likelihood (SML) method to estimate
the SV model. Subsequently, \citet{bib:sandmann} discussed the Monte Carlo likelihood estimation (MCL),
and how a very efficient MCL estimator can be obtained, while keeping the linear state-space form under
the classical inference. Their procedure first linearizes the SV model and then a better approximation of the
 observation equation error distribution is made using the MCL method. Further, \citet{bib:kim1998}
 used a normal mixture to approximate the observation distribution once the SV model is linearized. They also
 proposed a different estimation method based on Gibbs sampling under the Bayesian approach.

Another interesting approach is the method of moments (MM). Several MM estimators have
been introduced in the literature; for example, see  \citet{bib:taylor86} and \citet{bib:melino}. The latter used the
generalized method of moments (GMM) to estimate the SV model parameters. These estimators prevent
the problems related to model linearization and full likelihood function evaluation. However, they have poor
finite sample properties, and do not estimate the underlying volatility directly \citep{bib:BrRu2004}.

A Bayesian estimation approach to SV models using a Markov Chain Monte Carlo (MCMC) method and the full likelihood function
was developed by \citet{bib:jac1994}. Their extensive simulation experiments showed that the MCMC method
performs better than the QML and MM estimation techniques. Subsequently, \citet{bib:jac2004} introduced a
new version of this model to accommodate fat tails and correlated errors and \citet{bib:cappuccio2004} presented an interesting Skew-GED SV model.

The MCMC procedure requires large, computer-intensive simulations \citep{bib:BrRu2004} and its computational implementation
is not a simple task. Another problem is the dimensionality of the parameter space, once
the latent (log-volatility) and static parameters are simultaneously estimated using
the full posterior distribution, which is based on a full likelihood function, although it
is does not necessarily require the linearization of the SV model. Two alternatives
to the MCMC methods under the Bayesian perspective
are the particle filter \citep{bib:pitt1999,bib:lopes2011,bib:malik2011}
and Laplace \citep{bib:rue2009} approximations.

Several studies have examined using SV models from a Bayesian perspective, including those of \citet{bib:taylor1994}, \citet{bib:chib2002}, \citet{bib:yu2005}, \citet{bib:omori2007}, \citet{bib:raggi2006}, and \citet{bib:kaster2014}. Then,
\citet{bib:ferrante}, \citet{bib:vidoni}, and \citet{bib:davis2005} considered nonlinear and non-Gaussian state-space models. See also  \citet{bib:Wa1999}, \citet{bib:KnYu2002}, \citet{bib:feunou}, and \citet{bib:koopman2012}. A detailed review of SV models can be found in \citet{bib:BrRu2004}.

The family of non-Gaussian state-space
models (NGSSM) was proposed by \citet{bib:gamerman2013} and is an attractive alternative
to both the SV and GARCH models. These models have a dynamic
level associated with volatility and a multiplicative Beta evolution equation. This evolution
provides an exact marginal likelihood function and filtering and smoothing distributions. In spite of
the analytical tractably of this family of models, the
 evolution equation is a random walk in log-scale, and does not include drift (a quite restrictive). \citet{bib:pinho2015} presented
 several heavy-tailed distributions representing particular cases of the NGSSM family.
 In this class, \citet{bib:shepard1994} introduced local scale models, which were then generalized by \citet{bib:deschamps2011}.

 Several studies explore inferences based on stochastic volatility (SV) models, taking into
account the stylized facts of return data. The general problem with these models is that the latent parameters are high-dimensional, which makes it difficult to integrate out or to use high-dimensional numerical integration. Thus, inferences using these models require approximations using, for example, Markov Chain Monte Carlo or Laplace methods\citep{bib:jac1994,bib:jac2004,bib:omori2007,bib:kaster2014}. The GARCH model has been an attractive option among the users due to the difficult in obtaining the marginal likelihood of the SV model (its computational implementation) according to \citet{bib:frid1998}. Some SV models are expressed as a linear Gaussian state-space model, leading to an approximated marginal likelihood function and a marginal posterior distribution, which reduces the dimensionality of the problem. However, the observation disturbance is either Gaussian or requires approximations \citep{bib:harvey1994,bib:danielsson,bib:sandmann,bib:kim1998}. Other models are not linearized, and possess a marginal likelihood
that is approximated using Monte Carlo integration/importance sampling \citep{bib:davis2005}.

Thus, the main objective of this study is to develop a Bayesian GED-Gamma SV model for return data with a new sequential analysis procedure and an approximated marginal likelihood that is a product of the generalized Student's t-distributions and is evaluated directly, where the inferential procedure is fast and
easy to implement under the Bayesian approach. The latent states in our proposed GED-Gamma model are related across time through a stationary Gaussian evolution equation, and an analytical approximation is made for the prior distribution of the log-precision/volatility, without the need for model linearization. This also allows us to approximate the marginal likelihood function. Furthermore, the high-dimensional latent states are easily integrated out and sampled in blocks using a new approximated smoothing procedure that is introduced, enabling inferences to be made for these states.


 The main advantages of the employed method are its mathematical and computational simplicity, and its ability
to accommodate the stylized facts of return data and a stationary Gaussian evolution equation. This circumvents the
problem of high-dimensional latent states, without the need for model linearization.


Section 2 presents the GED-Gamma SV model. Then, Section 3 presents a simulation, and Section 4 provides
a case study of the proposed model using real return data. Finally, Section 5
concludes the paper, including an indication of potential areas for future research.

\section{GED-Gamma SV model}

Because of the stylized facts common to return data, we need to choose
more flexible models that allow for the use of non-Gaussian heavy-tailed skew
distributions \citep{bib:abad2014,bib:taylor86}. The GED is a non-Gaussian distribution
with the flexibility to capture heavy-tailed patterns, and is discussed in detail in
\citet{bib:box1992} and used in \citet{bib:nelson1991} and \citet{bib:deschamps2011}. Another possibility
is the skew-GED distribution that was used and motivated by \citet{bib:pinho2015} and \citet{bib:cappuccio2004}. However, we opt for a GED distribution that is a skew-GED distribution with the asymmetry parameter $\kappa=0$, as in \citet{bib:deschamps2011}. It is no difficult to extend the GED-Gamma SV model to other cases, as it will be shown in Subsection 2.3.

The GED-Gamma SV model, which is a composing of the GED distribution with precision distributed as a gamma distribution, for the return time series $\{y_t\}_{t=1}^{n}$ is defined
as follows:

\textbf{(A1)} \textbf{The observation equation} is
\begin{equation}
    p\left(y_{t}|\lambda_{t},\boldsymbol{\varphi}\right)=\frac{r\Gamma(3/r)^{1/2}}{2\Gamma(1/r)^{3/2}}\lambda_{t}^{1/r} \exp \left(-\lambda_{t}\psi(r)|y_{t}|^r\right),
\label{eq1}
\end{equation}
 for  $y_{t} \; \epsilon \; \Re$, where $\psi \left (r \right) = \left [\Gamma \left (3/r \right) / \Gamma \left (1/r \right) \right ]^{r/2}$, $\boldsymbol{\varphi}$ is a static parameter vector, the latent states $\lambda_{t}=h_t^{-1}$ (precision), and $h_t$ is the volatility at time $t$.  If $r=1$, it is the Laplace model, and if $r=2$, it is the normal model \citep{bib:deschamps2011}. We consider a correlation structure in the mean of the returns series, such that $y_t=(R_t-\mu_t)$, where $R_t$ is the usual return series and $\mu_t$ is the mean of the data.

 The model is fully specified by the following remaining assumptions:
\begin{itemize}
 \item \textbf{(A2)} \textbf{The prior distribution} is $\lambda_{t}|\mbox{\boldmath $Y$}_{t-1},\boldsymbol{\varphi}\sim\text{Gamma} (a_{t|t-1}, b_{t|t-1});$
    \item \textbf{(A3)} \textbf{The evolution equation} is $\ln(\lambda_{t})=-\alpha+\phi\ln(\lambda_{t-1})+\eta_{t}$, where $\eta_{t}\sim \text{i.i.d.  } \text{} N\left(0,\sigma_{\eta}^2\right)$, $\alpha\in\Re$, $\phi\in[0,1)$ and $\sigma_{\eta}^2>0$.
    \item \textbf{ The initial information} is $\lambda_{0}|Y_{0},\boldsymbol{\varphi}\sim \text{Gamma}(a_{0},b_{0})$, that is, $\ln(\lambda_{0})|Y_{0}\sim \text{Log-Gamma}(f_{0},q_{0}),$ where the mean is $f_0=\ln(a_{0})-\gamma(b_{0})$ and the variance $q_0=\gamma^{\prime}(a_{0})$. Then, $\gamma(\cdot)$ and $\gamma^{\prime}(\cdot)$ are the digamma and trigamma functions, respectively.
\end{itemize}

Note that $\mbox{\boldmath $Y$}_{t-1}=(Y_0,y_{1},\ldots,y_{t-1})^{'}$ is the information available up to time $t-1$. Furthermore, the evolution
equation $\textbf{(A3)}$ in terms of the volatility $h_t$ can be written as $\ln(h_t)=\alpha+\phi\ln(h_{t-1})+\eta_{t}^{\star},$ where $\eta_{t}^{\star}\sim N\left(0,\sigma_{\eta}^2\right)$ and $E(\varepsilon_t\eta_t^{\star})=0$, $\{\varepsilon_t\}$ is the disturbance term of the the observation equation.

\begin{figure}[htb]
    \centering
    \includegraphics[scale=0.8]{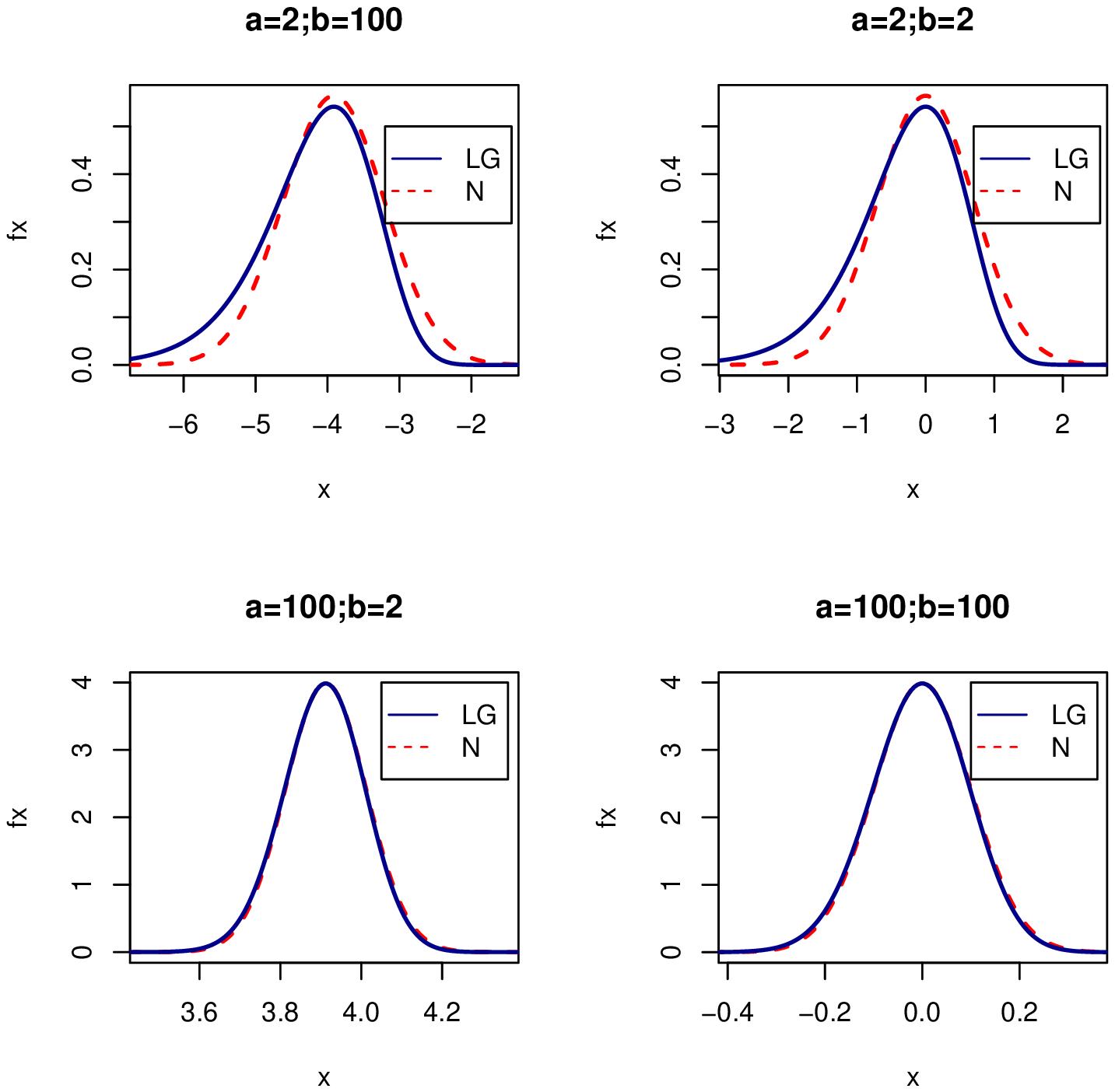}
    \caption{The log-gamma and normal distributions of the states for some values of the shape (a) and scale (b)
    parameters.}
    \label{approxlgn}
\end{figure}

Instead of approximations of the observation distribution, as in the QML, MCL, and MCMC \citep{bib:kim1998} methods, our approach approximates the distribution of the natural logarithm of the latent states, the log-precision, in terms of the two first moments, using an analytical approximation approach. Once the distribution of the natural logarithm of the latent states is a normal distribution or can be approximated by a normal, we can specify it in terms of its two first moments.
Figure \ref{approxlgn} shows a comparison of the log-gamma and normal
distributions for the states to illustrate and assess the quality of
the approximation in terms of two first moments. At the top, we have the shape parameter $(a)$ at 2 and the scale
parameter $(b)$ assuming the values 2 and 100 and a reasonable approximation of the
log-gamma distribution by the normal distribution. When the shape parameter is large,
the difference between the distributions become indistinguishable, because
of the central limit theorem. The values
of the parameters $a$ and $b$ were chosen based on the usual values of the
shape and scale parameters of the updated
distribution in our simulation experiments. This approach
is similar to that adopted in the dynamic generalized linear model (DGLM) \citep{bib:west1997}.

Hereafter, we present the proposed sequential analysis (inferential) procedure of this model, that is more similar to that of the Dynamic Linear Model (DLM) than the DGLM (see Figure \ref{seqproc}). This consists of the
one-step ahead predictive and filtering (or online) distributions of the latent states ${\boldmath\lambda}=\{\lambda_t\}_{t=1:n}$, and the one-step
ahead predictive distribution of the observations. If the model is defined as proposed in this section, we can use an approximation
of the state distribution to obtain the following results:
\begin{figure}[h]
    \centering
    \includegraphics[scale=0.45]{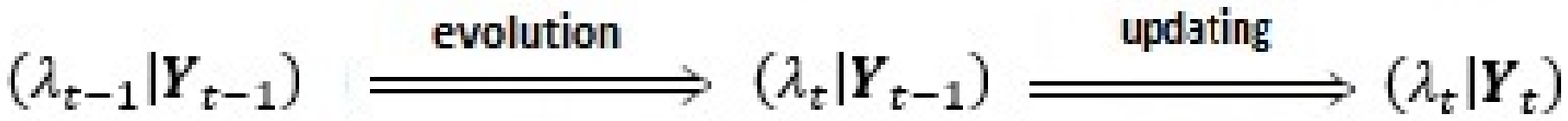}
    \caption{The sequential analysis procedure.}
    \label{seqproc}
\end{figure}

\textbf{Proposition 1.}
\begin{enumerate}
    \item The one-step ahead predictive (prior) distribution of the latent states at time $t$\\
    $\lambda_{t} |\mbox{\boldmath $Y$}_{t-1},\mbox{\boldmath $\varphi$}$ $\dot{\sim}\text{Gamma}(a_{t|t-1}, b_{t|t-1})$, where
\begin{eqnarray}
\label{eq33fg}
a_{t|t-1} &=& (\phi^2 a_{t-1}^{-1}+\sigma_{\eta}^2)^{-1},\\
b_{t|t-1} &=& \frac{\exp(\alpha)(a_{t-1}/b_{t-1})^{-\phi}}{(\phi^2 a_{t-1}^{^-1}+\sigma_{\eta}^2)}.
\end{eqnarray}
    \item The update or online (posterior) distribution at time $t$ $\lambda_{t}|{\textbf Y}_{t}, \mbox{\boldmath $\varphi$}\sim \text{Gamma}(a_{t},b_{t})$,
where
\begin{eqnarray}
\label{eq35fg1}
 \displaystyle a_{t} &=& a_{t|t-1}+1/r,\\
 \displaystyle b_{t} &=& b_{t|t-1} +\psi(r)|y_{t}|^r.
\end{eqnarray}

    \item The one-step ahead predictive distribution of the observations at time $t$ is given by
\begin{equation}
\label{eq310fg} p(y_t \vert \mbox{\boldmath $Y$}_{t-1},
\mbox{\boldmath $\varphi$}
)=\frac{\Gamma(1/r+a_{t|t-1})\frac{r\Gamma(3/r)^{1/2}}{2\Gamma(1/r)^{3/2}}
(b_{t|t-1})^{a_{t|t-1}}}{\Gamma(a_{t|t-1})[\psi(r)|y_{t}|^r
+b_{t|t-1}]^{1/r+a_{t|t-1}}}, \quad \\
y_{t}\in \Re,
\end{equation}
for $t=1,\ldots,n$, where $n$ is the number of observations of the time series and $\Gamma(\cdot)$ is the gamma function. This predictive distribution is the generalized Student's t-distribution with $2a_{t|t-1}$ degrees of freedom and if $r=2$, then it is Student's t-distribution \citep{bib:trianta2008}, an interesting feature of the proposed model.
\end{enumerate}
The proof of this proposition is given in Appendix I. The important distribution of $\lambda_{t} |\mbox{\boldmath $Y$}_{t-1},\mbox{\boldmath $\varphi$}$ in Part 1 of Proposition 1 preserves, in general, the mean of the distribution of $\lambda_{t-1} |\mbox{\boldmath $Y$}_{t-1},\mbox{\boldmath $\varphi$}$ and increases the variance.

The approximated marginal log-likelihood function, which is a product of the generalized Student's t-distributions, is given by
\begin{equation}
\label{eq311fg}
\begin{array}{l}
 \ln L(\mbox{\boldmath $\varphi$}; \mbox{\boldmath $Y$}_{n})=\ln\prod\limits_{t=1}^{n}p(y_t \vert \mbox{\boldmath $Y$}_{t-1}, \mbox{\boldmath $\varphi$})=\sum\limits_{t=1}^n  \ln
\Gamma (a_{t\mid t-1} +1/r)\\
-\ln \Gamma (a_{t\mid t-1})+
 a_{t|t-1}\ln b_{t|t-1}+\ln\left(\frac{r\Gamma(3/r)^{1/2}}{2\Gamma(1/r)^{3/2}}\right)\\-(1/r
 +a_{t|t-1})\ln[\psi(r)|y_{t}|^r+b_{t|t-1}],
\end{array}
\end{equation}
where $\mbox{\boldmath $\varphi$}$ is composed of $\alpha$, $\phi$, $\sigma_{\eta}^2$, and $r$; $\mbox{\boldmath $Y$}_{n}=(Y_0,y_{1},\ldots,y_{n})^{'} $ (when all information is available).





\subsection{Bayesian Inference}

Since the marginal posterior
distribution of parameter vector $\boldsymbol{\varphi}$ is not analytically tractable, a Bayesian inference for $\boldsymbol{\varphi}$ can be performed
using a MCMC \citep{bib:Ga2006} or an Adaptive Rejection Metropolis Sampling (ARMS) \citep{bib:Gi95} algorithm.
The marginal posterior distribution of $\mbox{\boldmath $\varphi$}$ is given by
\begin{eqnarray}
\label{eqibfg}
 p(\mbox{\boldmath $\varphi$}|\mbox{\boldmath $Y$}_{n})\propto L(\mbox{\boldmath $\varphi$};\mbox{\boldmath $Y$}_{n}) p(\mbox{\boldmath $\varphi$}),
\end{eqnarray}
where $L(\mbox{\boldmath $\varphi$};\mbox{\boldmath $Y$}_{n})$ is
the likelihood function defined in (\ref{eq311fg}), and
$p(\mbox{\boldmath $\varphi$})$ is the prior distribution of
$\mbox{\boldmath $\varphi$}$. In this work, independent proper uniform priors
are adopted for $\mbox{\boldmath $\varphi$}$, as in \citet{bib:gamerman2013} and \citet{bib:cappuccio2004}. The idea is to introduce vague uniform priors with a large variance, if we have no knowledge about the value of the parameters.
 However, other priors for the components of $\mbox{\boldmath $\varphi$}$ could be $\alpha\sim N(\mu_\alpha,\sigma_\alpha^2)$, $\frac{\phi+1}{2}\sim B(a_\phi,b_\phi)$, $\sigma^2\sim \text{InvGamma}(a_\sigma,b_\sigma)$,
and $r\sim \text{Gamma}(a_r,b_r)$ \citep[see]{bib:kaster2014}.

Once a sample $\boldsymbol{\varphi}^{(1)},\ldots, \boldsymbol{\varphi}^{(M)}$ is provided by the ARMS or MCMC algorithm,
the approximated posterior mean, median and percentiles can be calculated. The posterior mode can be obtained by maximizing function (\ref{eqibfg}). This task is typically performed numerically using a maximization algorithm, such as the Broyden\texttt{--}Fletcher\texttt{--}Goldfarb\texttt{--}Shanno (BFGS) and sequential quadratic programming (SQP) algorithms \citep{bib:avriel2003}. In general, $\mbox{\boldmath $\varphi$}$ may be re-parameterized in order to utilize these algorithms.









 An inference for the latent variables can be made using the output from the MCMC and ARMS algorithms.
Once a sample $\boldsymbol{\varphi}^{(1)}, ... , \boldsymbol{\varphi}^{(M)}$ is available,
the predictive, filtering or smoothed distributions of the latent states can be calculated in the following way. Note that
\begin{equation}
p(\lambda_{t+h} \vert \mbox{\boldmath $Y$}_{t} )= \int
 p(\lambda_{t+h} \vert \mbox{\boldmath $Y$}_{t}, \mbox{\boldmath $\varphi$} )\
p(\mbox{\boldmath $\varphi$}|\mbox{\boldmath $Y$}_{t}) \
d \mbox{\boldmath $\varphi$}. \label{intpred}
\end{equation}
Thus, the $h$-step-ahead predictive or filtering distributions can be approximated by
$$
\frac 1M \sum_{j=1}^M p(\lambda_{t+h} \vert \mbox{\boldmath $Y$}_{t}, \mbox{\boldmath $\varphi$}^{(j)} ),
$$
from which summaries such as means, variances, and credibility intervals can be obtained.
Since $p(\lambda_{t+h} \vert \mbox{\boldmath $Y$}_{t})$ is not available analytically, a draw
$\lambda_{t+h}^{(s)}$ from $p(\lambda_{t+h} \vert \mbox{\boldmath $Y$}_{t})$ can be obtained from (\ref{intpred}) by sampling
$\mbox{\boldmath $\varphi$}^{(s)}$ from $p(\mbox{\boldmath $\varphi$}|\mbox{\boldmath $Y$}_{n})$, and then sampling
$\lambda_{t+h}^{(s)}$ from $ p(\lambda_{t+h} \vert \mbox{\boldmath $Y$}_{t}, \mbox{\boldmath $\varphi$}^{(s)} )$. In addition, smoothing
procedures may be built, following \citet{bib:gamerman1991}. See also \citet{bib:migon2005}.

\subsection{Smoothing}

In order to infer the latent states $\lambda=(\lambda_1,\ldots,\lambda_n)^{'}$, we can utilize an approximated smoothed distribution
for $\ln(\lambda)$, and apply the inverse transformation. If the model is defined as proposed here, we can use the results of the sequential analysis to obtain the following smoothed distribution. The joint distribution of $\ln(\lambda) |\mbox{\boldmath $Y$}_{n},\boldsymbol{\varphi}$ has density
\begin{eqnarray}
\label{jointsmoot}
p(\ln(\lambda) | \boldsymbol{\varphi}, \mbox{\boldmath $Y$}_{n})
=p(\boldsymbol{\varphi}|\mbox{\boldmath $Y$}_{n}) p(\ln(\lambda_{n})|\boldsymbol{\varphi},\mbox{\boldmath $Y$}_{n})\prod_{t=1}^{n-1}p(\ln(\lambda_{t})|\ln(\lambda_{t+1}),\boldsymbol{\varphi},\mbox{\boldmath $Y$}_{t}).
\end{eqnarray}

\textbf{Proposition 2.}
 The distribution
\begin{equation}
\label{smoodist}
p(\ln(\lambda_{t})|\ln(\lambda_{t+1}),{\textbf Y}_{t}, \mbox{\boldmath $\varphi$})\doteq N(\mu_{t}^\star,\sigma_{t}^{2\star}),
\end{equation}
where $\sigma_{t}^{2\star}=\left(\frac{\phi^2}{\sigma_{\eta}^2}+\frac{1}{q_{t}}\right)^{-1}$, $\mu_{t}^{\star}=\sigma_{t}^{2\star}\left[\frac{\phi(\ln(\lambda_{t+1})+\alpha)}{\sigma_{\eta}^2}+\frac{f_{t}}{q_{t}}\right]$, $f_{t}=\ln(a_{t})-\gamma(b_{t})$ and $q_{t}=\gamma^{\prime}(a_{t})$, which depend on the shape and scale parameters of the filtering distribution of $\lambda_t$.
The proof of Proposition 2 is given in Appendix II.

The inference for the latent variables or states can be made using the output from the MCMC and ARMS algorithms.
Once a sample $\boldsymbol{\varphi}^{(1)}, ... , \boldsymbol{\varphi}^{(M)}$ is available,
posterior samples $\ln(\lambda)^{(1)} , ... , \ln(\lambda)^{(M)}$ from the latent variables are obtained according to the following procedure.

\textbf{Smoothing procedure:}
\begin{enumerate}

\item set $j=1$;

\item sample the static parameter $\boldsymbol{\varphi}^{(j)}$ from the MCMC or ARMS algorithm;

\item sample the set $\ln(\lambda)^{(j)}$ of latent variables from $p(\ln(\lambda)|\boldsymbol{\varphi}^{(j)},
\mbox{\boldmath $Y$}_{n})$ in (\ref{jointsmoot});

\item set $j \to j+1$ and return to 2, if $j \le M$; otherwise, stop.

\end{enumerate}

\subsection{Extensions of the GED-Gamma SV model}

The model, for the observations, in Equation (\ref{eq1}) can be generalized using a scale mixture for the observation disturbance to obtain other (skew) heavy-tailed distributions directly, such as
the (skew) Student's t-distribution \citep[see]{bib:nakajima2009,bib:gamerman2013}. If $\varepsilon_t^\star=\gamma_t^{-1/2}\varepsilon_t$ is the observation disturbance of the model, where $\gamma_t\sim\text{Gamma}(\nu/2,\nu/2)$ and $\varepsilon_t\sim\text{GED}(r=2,\mu=0,\sigma^2=1)$, $\varepsilon_t^\star$ will have a Student's t-distribution, with $\nu$ degrees of freedom \citep{bib:gamerman2013}.
Furthermore, other probability distributions may be considered for $\gamma_t$, leading to other (skew) heavy-tailed distributions for $\varepsilon_t^\star$. However, the (skew) GED specification in Equation (\ref{eq1}) leads to the one-step ahead predictive (skew) generalized Student's t-distribution for the observations (see Equation (\ref{eq310fg})) and the marginal likelihood that is a product of the (skew) generalized Student's t-distributions, which was also used by \citet{bib:wang2013} for modelling volatility data.

\section{A simulation}

We assess the performance of the proposed model using a Monte Carlo simulation, following the design of \citet{bib:sandmann,bib:jac1994}.

The values of $\mbox{\boldmath $\varphi$}=(\alpha,\phi,\sigma_{\eta}^2,r)^{T}$ are chosen in the following manner. First, we set the autoregressive parameter $\phi$ to 0.90, 0.95, and 0.98. Next, we take the values of $\sigma_{\eta}^2$ for each value of $\phi$, so as to ensure that the coefficient of variation (CV) $\exp(\frac{\sigma_{\eta}^2}{1-\phi^2})-1$ takes the values 10, 1, and 0.10. Then, we determine the values of $\alpha$, such that
the expected variance is equal to 0.0009. Finally, we set parameter $r$ to 2 and 1 and, thus, assume a Gaussian distribution ($\text{GED}(r=2)$) and Laplace distribution ($\text{GED}(r=1)$, the heavy-tailed case), respectively, for
the observation disturbance. 

 For each parameter setting, we generate 500 time series of length $n=500$ with normal and Laplace errors. We then estimate the proposed model and calculate
 the mean and mean squared error (MSE) of the posterior mode estimates. We adopt proper uniform priors for $\mbox{\boldmath $\varphi$}$. The prior distributions are $\phi\sim \text{Unif}(0,1)$, $\alpha\sim \text{Unif}(-10^3,10^3)$, $\sigma_{\eta}^2\sim \text{Unif}(0,10^3)$,
$r\sim \text{Unif}(0,10^3)$, and $\lambda_{0}|\mbox{ $Y$}_{0}\sim \text{Gamma}(0.001, 0.001)$, as in \citet{bib:gamerman2013}. Using the Bayesian approach, we use the Metropolis--Hastings (MCMC) with truncated normal proposed densities and the BFGS algorithms, implemented using Ox \citep{bib:ox}. We use two chains, 5,000 iterations of the MCMC algorithm, and a burn-in of 4,000 iterations. We perform simulations on a Pentium dual-core computer, with a 2.3 GHz processor and 4GB of RAM.

The parameter estimates of the GED-Gamma and normal-Gamma models are close to the true values for several settings with different coefficient variation values (see Table \ref{sim}) in the light-tailed case (normal errors). In general, the bias and MSE of the GED-Gamma model are close to the normal-Gamma model, which is the true model considered in this case.
The MSE values are small, and compete with other methods \citep[see]{bib:sandmann,bib:davis2005}. Among the four static parameters, $\alpha$ has the largest bias, in general. The bias for $\alpha$ and $\sigma_{\eta}^2$, with $CV=10$, is larger than with $CV=1$ and $0.1$. For $CV=10$, the bias of our method for $\alpha$ is slightly larger than those of the MCMC, QML \citep[see]{bib:sandmann}, IS, and AIS \citep[see]{bib:davis2005} methods. However, for $CV=0.1$, the estimates are not as biased as they are in \citet{bib:davis2005} and \citet{bib:sandmann}. For the heavy-tailed case (Laplace errors of the observation equation), clearly, the bias and MSE of the normal-Gamma model are larger than those in
the GED-Gamma model. This indicates a need for more flexible heavy-tailed models, such as the proposed GED-Gamma model in this work, and that ignoring flexible tails may lead you a poor scenario in terms of estimation.


\begin{table}
\centering
\caption{Comparison of static parameter estimates of the proposed GED-Gamma model, with different CV values and normal and Laplace errors, based on 500 replications.
For each parameter, the posterior mode estimate and mean square error are presented.}
 \label{sim}
\tiny
 \setlength{\tabcolsep}{3pt}

\def\arraystretch{1.0}
\begin{tabular}{c|cccccccccccc}
\hline\hline
           &                                                                                                    \multicolumn{ 12}{|c}{{\bf GED(r=2) (Normal) Errors }} \\
\hline
           &                                                                                                                      \multicolumn{ 12}{|c}{{\bf  CV=10 }} \\
\hline
           &    $\sigma_\eta^2$ &        $\phi$ &         $\mu$ &         $r$ &    $\sigma_\eta^2$ &        $\phi$ &         $\mu$ &         $r$ &    $\sigma_\eta^2$ &        $\phi$ &         $\mu$ &        $r$ \\
\hline
{\bf True} &      0.456 &      0.900 &     -0.821 &      1.000 &      0.234 &      0.950 &     -0.411 &      2.000 &      0.095 &      0.980 &     -0.164 &      2.000 \\
\hline
{\bf Normal} &      0.320 &      0.893 &     -1.019 &          - &      0.184 &      0.945 &     -0.530 &          - &      0.082 &      0.977 &     -0.223 &          - \\
 {\bf MSE} &      0.022 &      0.001 &      0.105 &          - &      0.004 &   2.93E-4 &      0.035 &          - &      0.001 &   8.45E-5 &      0.008 &          - \\
 {\bf GED} &      0.235 &      0.899 &     -0.876 &      1.838 &      0.153 &      0.945 &     -0.518 &      1.948 &      0.070 &      0.976 &     -0.226 &      1.998 \\
 {\bf MSE} &      0.056 &      0.001 &      0.085 &      0.098 &      0.010 &   3.42E-4 &      0.044 &      0.078 &      0.002 &   8.38E-5 &      0.011 &      0.080 \\
\hline
           &                                                                                                                         \multicolumn{ 12}{|c}{{\bf CV=1}} \\
\hline
{\bf True} &      0.018 &      0.900 &     -0.736 &      1.000 &      0.068 &      0.950 &     -0.368 &      2.000 &      0.028 &      0.980 &     -0.147 &      2.000 \\
\hline
{\bf Normal} &      0.017 &      0.920 &     -0.594 &          - &      0.059 &      0.952 &     -0.373 &          - &      0.025 &      0.981 &     -0.151 &          - \\
 {\bf MSE} &      0.000 &      0.001 &      0.033 &          - &      0.001 &   1.55E-4 &      0.009 &          - &   1.41E-4 &   3.22E-5 &      0.002 &          - \\
 {\bf GED} &      0.019 &      0.896 &     -0.799 &      2.053 &      0.055 &      0.946 &     -0.424 &      2.024 &      0.022 &      0.978 &     -0.171 &      2.033 \\
 {\bf MSE} &      0.002 &      0.001 &      0.160 &      0.063 &      0.001 &   2.26E-4 &      0.022 &      0.081 &   1.76E-4 &   3.72E-5 &      0.003 &      0.068 \\
\hline
           &                                                                                                                       \multicolumn{ 12}{|c}{{\bf CV=0.1}} \\
\hline
{\bf True} &      0.132 &      0.900 &     -0.706 &      1.000 &      0.009 &      0.950 &     -0.353 &      2.000 &      0.004 &      0.980 &     -0.141 &      2.000 \\
\hline
{\bf Normal} &      0.120 &      0.901 &     -0.746 &          - &      0.008 &      0.956 &     -0.315 &          - &      0.003 &      0.981 &     -0.134 &          - \\
 {\bf MSE} &      0.002 &      0.001 &      0.061 &          - &   4.28E-5 &   8.70E-5 &      0.004 &          - &   9.11E-6 &   8.79E-6 &   3.81E-4 &          - \\
 {\bf GED} &      0.102 &      0.895 &     -0.792 &      2.009 &      0.008 &      0.949 &     -0.373 &      2.045 &      0.003 &      0.979 &     -0.148 &      2.039 \\
 {\bf MSE} &      0.004 &      0.001 &      0.085 &      0.091 &   6.47E-5 &   8.66E-5 &      0.007 &      0.049 &   7.82E-6 &   9.80E-6 &      0.001 &      0.060 \\
\hline
           &                                                                                                              \multicolumn{ 12}{|c}{{\bf GED(r=1) (Laplace) Errors}} \\
\hline
           &                                                                                                                      \multicolumn{ 12}{|c}{{\bf  CV=10 }} \\
\hline
          &    $\sigma_\eta^2$ &        $\phi$ &         $\mu$ &         $r$ &    $\sigma_\eta^2$ &        $\phi$ &         $\mu$ &         $r$ &    $\sigma_\eta^2$ &        $\phi$ &         $\mu$ &        $r$ \\
\hline
{\bf True} &      0.456 &      0.900 &     -0.821 &      1.000 &      0.234 &      0.950 &     -0.411 &      1.000 &      0.095 &      0.980 &     -0.164 &      1.000 \\
\hline
{\bf Normal} &      1.138 &      0.869 &     -2.869 &          - &      0.753 &      0.927 &     -1.655 &          - &      0.427 &      0.967 &     -0.771 &          - \\
 {\bf MSE} &      0.489 &      0.002 &      4.554 &          - &      0.283 &      0.001 &      1.719 &          - &      0.118 &   3.58E-4 &      0.422 &          - \\
 {\bf GED} &      0.095 &      0.892 &     -0.486 &      1.001 &      0.052 &      0.944 &     -0.257 &      1.008 &      0.022 &      0.976 &     -0.107 &      1.003 \\
 {\bf MSE} &      0.132 &      0.001 &      0.140 &      0.016 &      0.034 &   3.97E-4 &      0.034 &      0.013 &      0.006 &   1.07E-4 &      0.005 &      0.010 \\
\hline
           &                                                                                                                         \multicolumn{ 12}{|c}{{\bf CV=1}} \\
\hline
{\bf True} &      0.018 &      0.900 &     -0.736 &      1.000 &      0.068 &      0.950 &     -0.368 &      1.000 &      0.028 &      0.980 &     -0.147 &      1.000 \\
\hline
{\bf Normal} &      0.303 &      0.851 &     -2.349 &          - &      0.400 &      0.922 &     -1.359 &          - &      0.226 &      0.969 &     -0.563 &          - \\
 {\bf MSE} &      0.099 &      0.005 &      3.259 &          - &      0.120 &      0.001 &      1.135 &          - &      0.044 &   2.44E-4 &      0.203 &          - \\
 {\bf GED} &      0.025 &      0.850 &     -0.642 &      1.047 &      0.016 &      0.945 &     -0.210 &      1.015 &      0.006 &      0.977 &     -0.087 &      1.014 \\
 {\bf MSE} &      0.009 &      0.038 &      1.096 &      0.018 &      0.003 &   2.94E-4 &      0.030 &      0.011 &   4.82E-4 &   6.89E-5 &      0.005 &      0.009 \\
\hline
           &                                                                                                                       \multicolumn{ 12}{|c}{{\bf CV=0.1}} \\
\hline
{\bf True} &      0.132 &      0.900 &     -0.706 &      1.000 &      0.009 &      0.950 &     -0.353 &      1.000 &      0.004 &      0.980 &     -0.141 &      1.000 \\
\hline
{\bf Normal} &      0.625 & 0.845 &     -2.540 &          - &      0.182 &      0.929 &     -1.095 &          - &      0.094 &      0.975 &     -0.395 &          - \\
 {\bf MSE} &      0.260 &  0.005 &      3.769 &          - &      0.040 &      0.001 &      0.715 &          - &      0.012 &   7.90E-5 &      0.079 &          - \\
 {\bf GED} &      0.037 &      0.878 &     -0.457 &      1.021 &      0.004 &      0.947 &     -0.195 &      1.023 &      0.001 &      0.979 &     -0.077 &      1.032 \\
 {\bf MSE} &      0.010 &      0.011 &      0.228 &      0.015 &   9.46E-4 &   6.38E-4 &      0.056 &      0.011 &   6.67E-6&   1.68E-5 &      0.004 &      0.008 \\
\hline\hline
\end{tabular}

\end{table}

\section{A case study with return data}


This case study uses the daily return data a Petrobr\'{a}s (a Brazilian company) asset and the pound/dollar exchange rate. The first is
for the period 02/01/2001 to 06/02/2015 (3546 observations), and the second is for the period
10/01/1981 to 06/28/1985 (946 observations). The data can be found at the Yahoo finance website, and the second data set is also available in \citet{bib:durbin2001}. Here,
the return series at time $t$ is defined as
$y_t=R_t=100\ln\left(\frac{P_{t}}{P_{t-1}}\right)$, centered around the sample mean, where $P_{t}$ is the daily closing
spot price. For the second data set, $P_t$ represents the daily closing exchange rate. Data irregularity due to holidays and weekends
is ignored. We perform our case study using Ox \citep{bib:ox} installed on a Pentium dual-core
computer, with a 2.3 GHz processor and 4GB of RAM. The codes are available upon authors request.

\begin{figure}[h]
    \centering
    \includegraphics[scale=0.6]{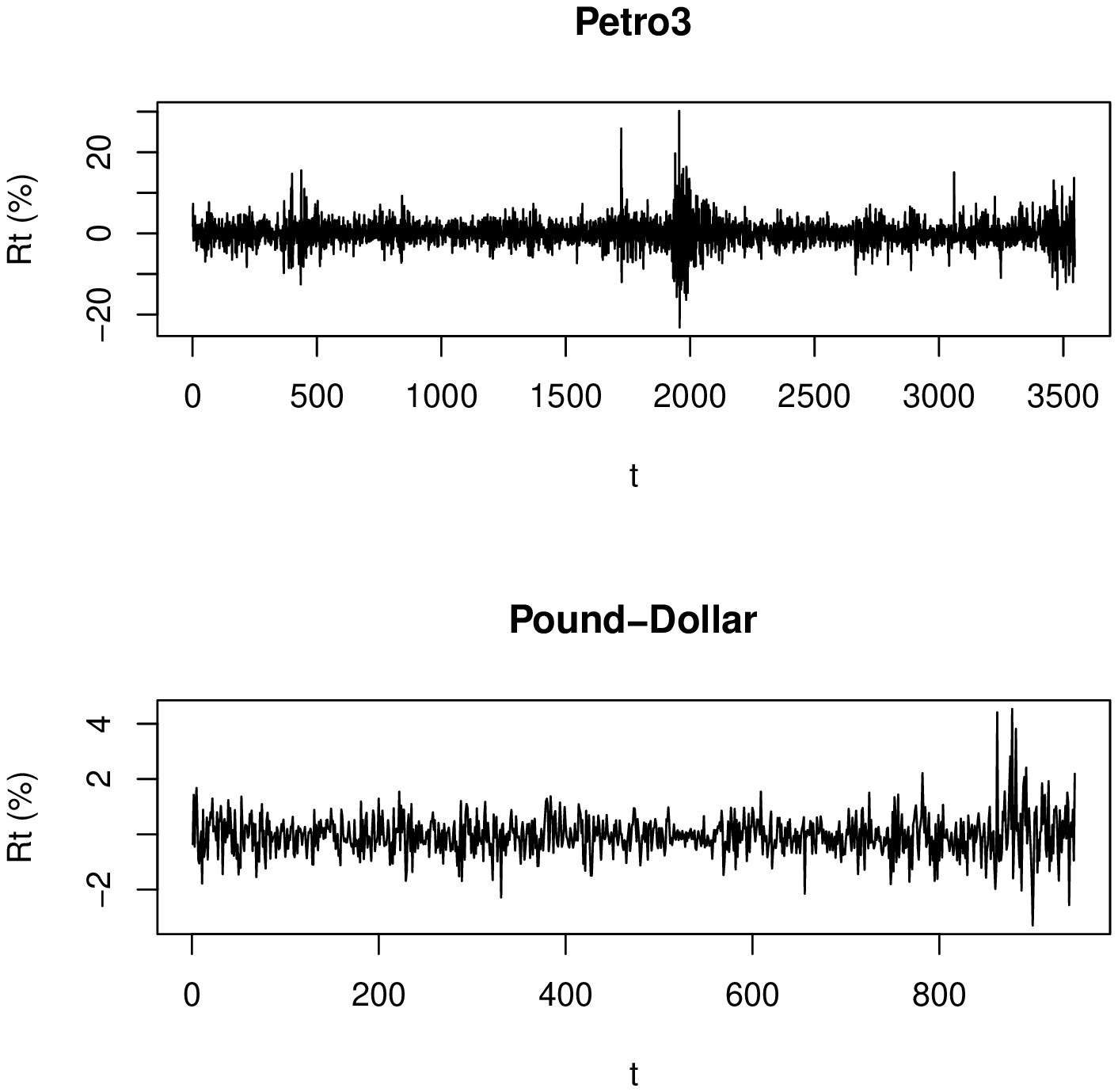}
    \caption{The Petrobr\'{a}s and pound/dollar return series.}
    \label{logretpetro}
\end{figure}

Figure \ref{logretpetro} presents the time
series plots of the Petrobr\'{a}s and pound/dollar returns. The Pound/Dollar return data set was
analyzed by \citet{bib:harvey1994} and then reanalyzed by \citet{bib:davis2005}. A distinctive feature of
financial time series is that they usually present nonconstant
variance or volatility (see Figure \ref{logretpetro}). Descriptive statistics are shown in Table \ref{descstat}. The Petrobr\'{a}s return series
presents an excess of kurtosis compared to that of the pound/dollar returns. Both series have a slight positive skewness.

\begin{table}
\centering
\caption{Descriptive statistics for the return series.}
 \label{descstat}
\footnotesize
 \setlength{\tabcolsep}{3pt}
\def\arraystretch{1.0}

\begin{tabular}{c|c|c}
\hline\hline

           & {\bf Petrobr\'{a}s  } & {\bf Pound/Dollar} \\
\hline
{\bf No. of obs} &       3546 &        946 \\
{\bf Mean} &     0.0000 &     0.0000 \\
{\bf Median} &   0.003662 &    -0.0104 \\
{\bf Std. dev.} &      3.059 &      0.711 \\
{\bf Skewness} &      0.340 &      0.604 \\
{\bf Kurtosis} &      11.54 &      7.862 \\
{\bf P-value (Normality test)} &   0.000   &  0.000    \\
\hline\hline
\end{tabular}
\end{table}

The proposed GED-Gamma model is fitted to
the return data of these assets  using the Bayesian approach and the Metropolis--Hastings (MCMC)
and BFGS algorithms, implemented in Ox \citep{bib:ox}. For our model, we adopt independent proper uniform
priors for $\mbox{\boldmath $\varphi$}$. The prior distributions for the parameters are $\phi\sim \text{Unif}(0,1)$, $\alpha\sim \text{Unif}(-10^3,10^3)$, $\sigma_{\eta}^2\sim \text{Unif}(0,10^3)$,
$r\sim \text{Unif}(0,10^3)$, and $\lambda_{0}|\mbox{$Y$}_{0}\sim \text{Gamma}(0.001, 0.001)$. We used two chains, 5,000 iterations of the MCMC algorithm, and a burn-in of 4,000 iterations.

\begin{figure}[h]
    \centering
    \includegraphics[scale=0.5]{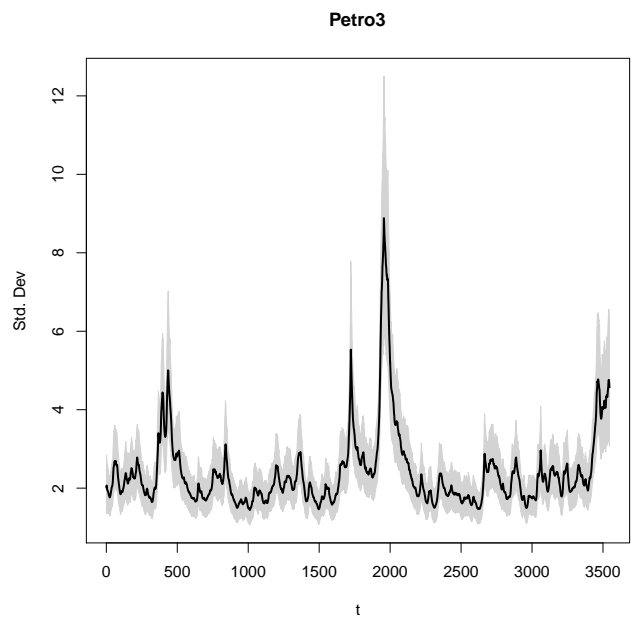}
    \caption{The
    smoothed mean volatility obtained using the proposed GED-Gamma volatility model for the Petrobr\'{a}s returns. The grey area indicates the 95\% credibility intervals.}
    \label{filtvolpet3}
\end{figure}

\begin{figure}[h]
    \centering
    \includegraphics[scale=0.5]{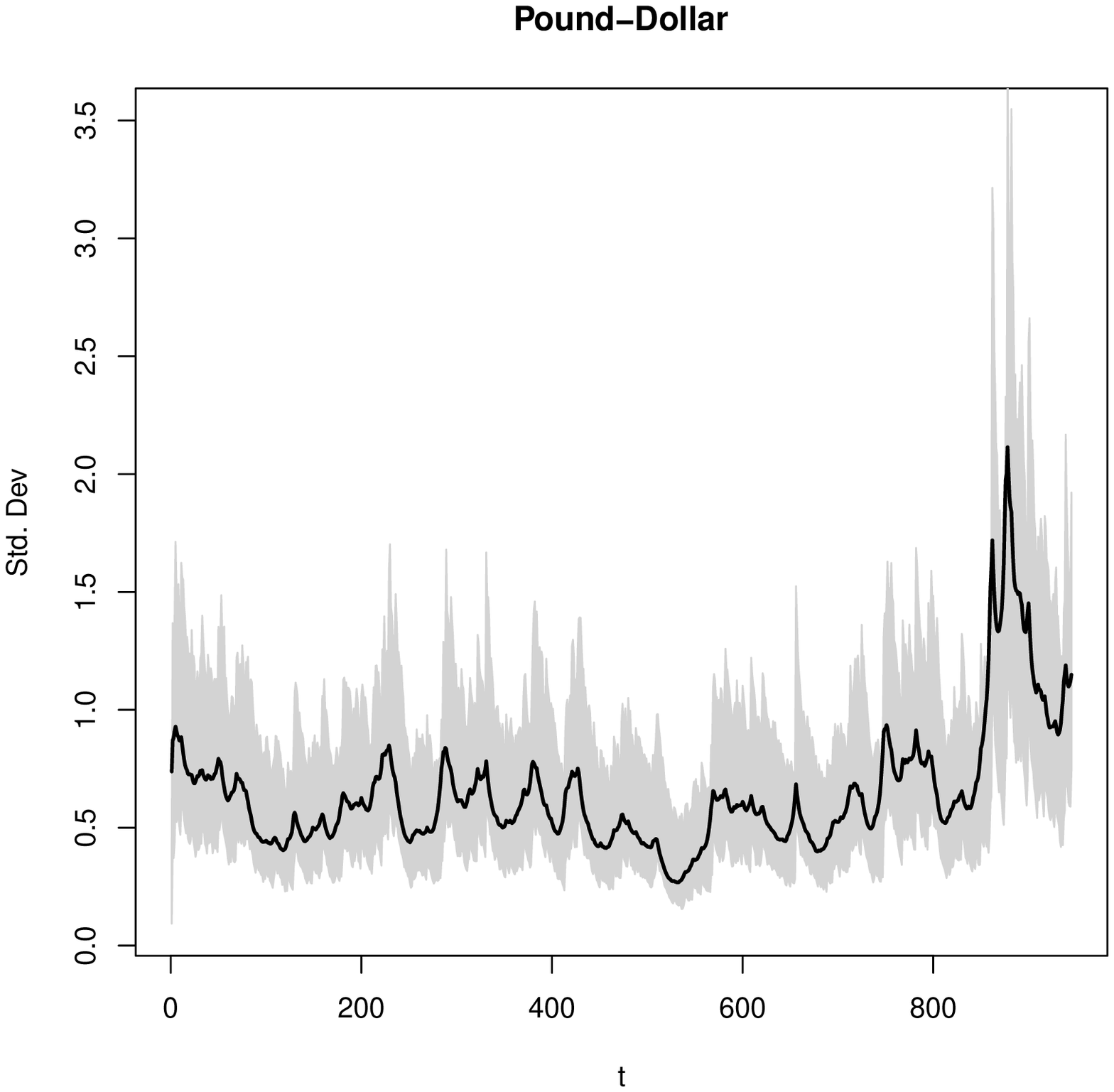}
    \caption{The
    smoothed mean volatility obtained using the proposed GED-Gamma volatility model for the pound/dollar exchange rate. The grey area indicates the 95\% credibility intervals.}
    \label{filtvolpound}
\end{figure}

  Table \ref{maxlogvero} shows the log-likelihood value and the Bayes factor used to evaluate the model fit.
For the Petrobr\'{a}s returns, the results show that our GED-Gamma model outperforms the normal-Gamma model. There is strong evidence
in favor of the GED-Gamma model ($BF=0.0067$), which allows a more flexible heavy-tailed distribution for the observation disturbance and this was also indicated for the heavy-tailed case in the simulation. For pound/dollar returns, the results are similar, with a slight, but negligible preference by the normal model, according to the Bayes factor. The parameter estimates for the GED-Gamma and normal-Gamma models are shown in Table \ref{estapp}. A residual analysis shows no strong violation of the proposed model assumptions. Figures \ref{filtvolpet3} and \ref{filtvolpound} show the smoothed volatility for the two assets, using the procedure described in Subsection 2.2. The estimated volatility follows the volatility pattern of the return series well, and presents peaks that correspond with crisis periods.

\begin{table}[htb]
\centering
\caption{Values of the log-likelihood and Bayes factors for the proposed GED-Gamma and normal-Gamma models fitted to the Petrobr\'{a}s and pound/dollar returns.}
 \label{maxlogvero}
\footnotesize
\centering
 \setlength{\tabcolsep}{6pt}
\def\arraystretch{1.0}
\begin{tabular}{c|c|c|c}
\hline
\hline
           &            & \multicolumn{ 2}{|c}{{\bf Assets}} \\
\hline
{\bf Methods} & {\bf Criterion} & {\bf Pound/Dollar} & {\bf Petro3} \\
\hline
 {\bf GED} &     LogLik$^\dag$ &    -925.74 &   -8390.52 \\
           &    MLogLik$^{\star\star}$ &    -928.94 &   -8393.00 \\
\hline
{\bf Normal$^\star$} &     LogLik$^\dag$ &    -925.97 &   -8396.45\\ 
           &    MLogLik$^{\star\star}$ &    -928.03 &   -8398.00 \\
           &         BF &       2.48 &        0.0067 \\
\hline
\hline
\end{tabular}

\textbf{Note}: Bayes factors against the proposed GED-Gamma model. $^\dag$at the posterior mode; $^\star$the normal model is the proposed GED-Gamma model, with $r=2$; $^{\star\star}$marginal log-likelihood after integrating out the parameters.
\end{table}

\subsection{Comparisons of the competing models}

In this section, we compare the GED-Gamma model to other
models, including the Kim,
Shephard, and Chib SV model (SV-KSC) \citep{bib:kim1998}, and the Bayesian GARCH(1,1) model
with Student's t-disturbances \citep{bib:Ardia2008} (t-GARCH(1,1)). The former is implemented using
Ox, by \citet{bib:pelagatti2011}, and
latter using the R package 'bayesGARCH' \citep{bib:ardia2015}.


For the SV-KSC approach, the simplest SV model is linearized, so $\log(y_t^2)$ is used. An approximation
for the $\log \chi_1^2$ distribution is performed using a seven-component Gaussian mixture. Therefore,
conditional on latent indicator variables $w_t\in\{1,\ldots,7\}$, $t=1,\ldots,n$, the SV-KSC model is given by
\begin{eqnarray}
\label{KSC-SV}
  \log y_t^2 - \mu_{w_t}&=& h_t +\log \varepsilon_t^2,  \varepsilon_t\sim N(0,\sigma_{w_t}^2), \\
  h_t &=& \alpha+\phi h_{t-1}+\eta_t,  \eta_t\sim N(0,\sigma_\eta^2),
\end{eqnarray}
$t=1,\ldots,n$. We specify the independent noninformative
priors for the model parameters given by $p(\alpha,\phi)\propto p_{N_2}(\alpha,\phi)$ and
$p(\sigma_\eta^2)\propto p_{IG}(\sigma_\eta^2),$ where IG denotes the inverse gamma distribution. The hyperparameters of
the priors are the package default values.

Another competing model is the GARCH(1,1) model with Student's t-innovations \citep{bib:Ardia2008}, given as:
\begin{equation}\label{t-garch}
y_t= \left( \frac{(\nu-2)h_t}{\nu}\right)^{-1/2}\varepsilon_t,
\end{equation}
$t=1,\ldots,n$, where the conditional variance equation is $h_t=\alpha_0+\alpha_1y_{t-1}^2+\beta h_{t-1},$
$\alpha_0>0$, $\alpha_1\geq0$ and $\beta\geq0$ to ensure a positive conditional variance and, finally, $\varepsilon_t\overset{i.i.d.}{\sim} \text{Student-t} (\nu)$.
The independent prior used in the t-GARCH(1,1) model is:
$p(\alpha_0)\propto p_N(\alpha_0;0,1000)I_{[\alpha_0>0]}$, $p(\alpha_1)\propto p_N(\alpha_1;0,2000)I_{[\alpha_1>0]}$, $p(\beta)\propto p_N(\beta;0,1000)I_{[\beta>0]}$
and $p(\nu)=0.01\exp(-0.01(\nu-2))I_{[\nu>2]}$.

For the t-GARCH(1,1) and the SV-KSC models, we utilized one chain, 12,000 iterations of the MCMC algorithm, and a burn-in
of 10,000 iterations for the Bayesian methods. The convergence of the chain was
checked using methods, such as graphs.



Table \ref{estapp} presents the static parameter estimates of the proposed approach, the SV-KSC, and t-GARCH(1,1) methods.
The estimates of our proposed model with normal innovations are very close to those of the SV-KSC method, with normal innovations,
in most of the cases for the two assets. The interval estimates (Table \ref{estapp}) show that the needed
of a more flexible model than the normal-Gamma, as the GED-Gamma model.




For the pound/dollar returns, our parameter estimates (Table \ref{estapp}) are very close to those obtained by \citet[p.397]{bib:davis2005} and \citet[p.236]{bib:durbin2001}. The QML
estimates from \citet{bib:harvey1994} are similar to ours, but mainly in the normal-Gamma case. Furthermore, the parameter estimates of
our model are very similar to those of the SV-KSC (Table \ref{estapp}). The parameters $\alpha_0$, $\alpha_1$, $\beta$,
and $\nu$ belong to the t-GARCH(1,1). The estimate of $\nu$ indicates a heavy-tailed pattern for the two return series.

\begin{table}[htb]
\centering
\caption{Static parameter estimates of the models fitted to the Petrobr\'{a}s and pound/dollar returns.}
 \label{estapp}
\scriptsize
 \setlength{\tabcolsep}{6pt}
\def\arraystretch{1.0}

\begin{tabular}{c|c|c|c|c}
\hline
\hline
                       &               \multicolumn{ 4}{|c}{{\bf Methods}} \\
\hline
                       & {\bf GED$^\S$} & {\bf Normal$^{\S\S}$}  &  {\bf SV-KSC} &  {\bf t-GARCH(1,1)}\\ 
                       & {\bf P. Mode } & {\bf P. Mode  }   &  {\bf P. Mean} &  {\bf P. Mean}\\ 
               {\bf }                 & {\bf   (P. Mean)} & {\bf (P. Mean)  }  & {\bf   } & \\ 
\hline
 {\bf Estimates} &         \multicolumn{ 4}{|c}{{\bf Petrobr\'{a}s}}      \\         

\hline
 {\bf $\hat{\sigma}_{\eta}^2$} &     0.011 (0.012) &   0.019 (0.021)  &          0.025 & - \\ 


      {\bf CI} & [0.0065;0.0226] & [0.0140;0.0310]  & [0.0161;0.0375]  & -\\ 
      {\bf $\hat{\phi}$} &     0.985 (0.983) &      0.980 (0.978)  &           0.980 & -\\  


       {\bf CI} & [0.9685;0.9929] &   [0.9670;0.9868]  &  [0.9680;0.9889] & -\\ 
      {\bf $\hat{\alpha}$} &     0.019 (0.022)  &  0.028 (0.031)   &            0.037 & -\\ 


       {\bf CI} & [0.0083;0.0426]  & [0.0167;0.0490]   &  [0.0198;0.0588] & -\\ 
        {\bf $\hat{r}$} &      1.725 (1.744) &  -      &   -       & -\\ 


       {\bf CI} & [1.5972;1.9041] & - &        -   & -\\   







   {\bf $\hat{\alpha}_0$} &      -  &   -      &    -        & 0.205 \\

             {\bf CI} & - &        -       &   -    & [0.13498;0.2900] \\


      {\bf $\hat{\alpha}_1$} &   - &   -      &    -        & 0.082 \\

             {\bf CI} & - &        -       &   -    & [0.06468;0.1023] \\

     {\bf $\hat{\beta}$} &    - &   -      &    -        & 0.894 \\
               {\bf CI} & - &        -       &   -    & [0.87091;0.9154] \\

     {\bf $\hat{\nu}$} &     - &   -      &    -        & 7.289 \\

           {\bf CI} & - &        -       &   -    & [5.7441;9.0390] \\

\hline
 {\bf Estimates  } &    \multicolumn{ 4}{|c}{{\bf Pound/Dollar}}  \\ 
\hline
 {\bf $\hat{\sigma}_{\eta}^2$} &     0.019 (0.047) &  0.025 (0.042) &           0.036 & -\\ 


       {\bf CI} & [0.0139; 0.0999] &  [0.0159;0.0880]  & [0.0155;0.0681] & -\\ 
      {\bf $\hat{\phi}$} &     0.978 (0.956) &   0.974 (0.957) &          0.969 & -\\ 


  {\bf CI} & [0.9107;0.9847] &  [0.9054;0.9878] & [0.9376;0.9917]  & -\\ 
      {\bf $\hat{\alpha}$} &     -0.028 (-0.063)&   -0.036 (-0.059) &      -0.031  & -\\ 


      {\bf CI} & [-0.1755;-0.0193] &  [-0.1288;-0.0191]  & [-0.0669;-0.0057] & -\\
       {\bf $\hat{r}$} &      1.884 (2.049)&   -      &    -        & - \\


      {\bf CI} & [1.7097;2.4793] &        -       &   -    & - \\ 

   {\bf $\hat{\alpha}_0$} &      -  &   -      &    -        & 0.031 \\

             {\bf CI} & - &        -       &   -    & [0.0148;0.0557] \\


      {\bf $\hat{\alpha}_1$} &   - &   -      &    -        & 0.138 \\

             {\bf CI} & - &        -       &   -    & [0.08970;0.20394] \\

     {\bf $\hat{\beta}$} &    - &   -      &    -        & 0.810 \\
               {\bf CI} & - &        -       &   -    & [0.7323;0.8664] \\

     {\bf $\hat{\nu}$} &     - &   -      &    -        & 8.635\\

           {\bf CI} & - &        -       &   -    & [4.9359;13.7402] \\
\hline
\hline
\end{tabular}

{\bf Note}: $^\S$The proposed GED-Gamma model; $^{\S\S}$the proposed GED-Gamma model, with $r=2$ (normal case); CI: 95\% percentile credibility interval.

\end{table}

For the in-sample analysis of the GED-Gamma model, we use the smoothed mean of volatility, calculated using the
smoothing procedure of Subsection 2.2. For the SV-KSC and t-GARCH(1,1) models, we use the posterior mean of the volatility. The square of the log-return is used
as a proxy for the true unobserved
volatility $\sigma_t^2$ \citep{bib:Ba2012}. Thus, the square root of the mean squared error, $\text{SRMSE}=\sqrt[]{\frac{\sum\limits_{t=1}^n(y_{t}^2 - \hat{\sigma}_{t}^2)^{2}}{n}}$, and mean absolute error, $\text{MAE}=\frac{\sum\limits_{t=1}^n|y_{t}^2 - \hat{\sigma}_{t}^2|}{n}$, are used to compare the models. For the in-sample analysis of the pound/dollar returns, the GED-Gamma
model has the smallest SRMSE value (Table \ref{inofsample}). For the Petrobr\'{a}s return series, the smallest MAE value.
In most cases, the GED-Gamma model is
the best or second best of the competing models, indicating that it performs well in terms of fit.

\begin{table}[htb]
\centering
\caption{The SRMSE and MAE of in-sample estimation of the volatility of the proposed GED-Gamma, SV-KSC and t-GARCH models
 fitted to the Petrobr\'{a}s and pound/dollar returns.}
 \label{inofsample}
\small
 \setlength{\tabcolsep}{5pt}
\def\arraystretch{0.9}
\begin{tabular}{c|c|c|c|c}
\hline
\hline
{\bf Assets} &            & {\bf GED Model} & {\bf SV-KSC} & {\bf GARCH-t} \\
\hline
{\bf Petrobr\'{a}s} & {\bf SRMSE} &      27.23  (2)&      25.89 (1)&      30.34 (3)\\
    {\bf } &  {\bf MAE} &       8.36 (1)&       8.87 (3)&       8.53 (2)\\
\hline
{\bf Pound/Dollar} & {\bf SRMSE} &       1.15 (1)&       1.25 (2)&       1.34 (3)\\
           &  {\bf MAE} &       0.48 (2)&       0.59 (3)&       0.47 (1)\\
\hline
\hline
\end{tabular}

\textbf{Note:} The numbers in parentheses denote the ranking among the competing models.
\end{table}

\subsubsection{Out-of-sample forecast comparisons}

For the out-of-sample analysis, a direct comparison
of volatility forecasts is adopted
using the square of the log-return as a proxy for the true unobserved
volatility $\sigma_{t+1}^2$ \citep{bib:Ba2012}. For the proposed GED-Gamma model, the one-step ahead forecast volatility
$\hat{\sigma}_{t+1}^2$ is calculated using the distribution of Item 1 on page 5.
Under the Bayesian approach, the SRMSE and MAE are computed using the one-step ahead
forecast $\hat{\sigma}_{t+1}^2$ of the competing models, leaving the out last five observations, then the last four observations
out, and so on, until the last observation is left out. Finally, the SRMSE and MAE are
computed as
$\sqrt[]{\frac{\sum\limits_{k=1}^5(y_{t+k}^2 - \hat{\sigma}_{t+k}^2)^{2}}{5}}$ and $\frac{\sum\limits_{k=1}^5|y_{t+k}^2 - \hat{\sigma}_{t+k}^2|}{5}$, respectively, where
the index $k$ varies over the last five observations.

Table \ref{outofsample} presents the SRMSE and MAE of one-step ahead forecasts of the proposed GED-Gamma, SV-KSC, and t-GARCH(1,1) models.  For the out-of-sample
analysis of the pound/dollar returns, the SRMSE and MAE values of the GED-Gamma and SV-KSC models are similar, while the MAE and MSE of the GED-Gamma model are smaller than those of the SV-KSC and t-GARCH(1,1) models for the Petrobr\'{a}s return series. In most cases, the GED-Gamma model is the best or second best of the three competing models. This indicates that the proposed GED-Gamma model is also a good option in terms of prediction.

\begin{table}[htb]
\centering
\caption{The SRMSE and MAE of the one-step ahead forecasts for the volatility of the proposed GED-Gamma, SV-KSC and t-GARCH models
 fitted to the Petrobr\'{a}s and pound/dollar returns.}
 \label{outofsample}
\small
 \setlength{\tabcolsep}{5pt}
\def\arraystretch{0.9}
\begin{tabular}{c|c|c|c|c}
\hline
\hline
{\bf Assets} &            & {\bf GED Model} & {\bf SV-KSC} & {\bf t-GARCH(1,1)} \\
\hline
{\bf Petrobr\'{a}s} & {\bf SRMSE} &      80.77 (1)&      86.58 (2)&      86.78 (3)\\
    {\bf } &  {\bf MAE} &      56.57 (1)&      57.05 (2)&      59.45 (3)\\
\hline
{\bf Pound/Dollar} & {\bf SRMSE} &       1.88 (2)&       1.84  (1)&       2.06 (3)\\
           &  {\bf MAE} &       1.27 (2)&       1.10 (1)&       1.29 (3)\\
\hline
\hline
\end{tabular}

\textbf{Note:} The numbers in parentheses denote the ranking among the competing models.
\end{table}

\section{Conclusion}

In this study, we introduced a GED-Gamma SV model for return data with an approximated expression for the marginal likelihood, which can be evaluated directly, under the Bayesian approach. Using the model, we propose
new sequential analysis and smoothing procedures and a marginal likelihood that is a product of the generalized Student's t-distributions based on an analytical approximation for the distribution of the latent states. 
The main
advantages of the proposed method are its mathematical and computational simplicity and its ability
to accommodate the stylized facts of return data and a stationary Gaussian evolution equation, circumventing the
problem of the high-dimensional latent states.
There is no
need to linearize the model; that is,
the data scale is not changed and is free from approximation of the
observation distribution. Non-Gaussian, heavy-tailed skew
distributions for the observations are naturally accommodated. Beyond of the approximated sequential analysis
procedure, the smoothing procedure is provided. Another interesting feature
is the availability of the one-step ahead predictive distribution, which is the generalized Student-t distribution.

A limitation of the model is the use of
approximations for the distribution of the latent states in terms of the two first moments, because
it was developed as a DGLM \citep{bib:west1997,bib:souza2018}. The quality of this approximation depends on the
quality of the normal approximation to the log-gamma prior distribution of the latent states. The DGLM has a
dynamic structure in the mean of the data, which here is volatility. Both methods preserve
the sequential analysis of the data.


Our approach performed well in the parameter estimation
of the GED-Gamma SV model using the posterior mode, mean and quantiles
 under the Bayesian perspective. The empirical results are competitive compared to other methods
 in the literature in terms of fit and prediction. Thus, we achieved our primary objective of introducing a Bayesian
 GED-Gamma SV model that can be implemented in a fast and easy way, and that is free of approximations
 for the observation equation. The results and the proposed procedures of this
 study are also useful to the closely related time series model. For example, the dynamic
 linear models proposed by \citet{bib:west1997}, for normal observations
with time varying means and variances, allows a stationary evolution equation for the volatility. Our results
can also be used in the model of \citet{bib:nakajima2009}, without the need to linearize the model for
the volatility sampling.


Future works could include a study of other distributions for the observation
equation (especially skew distributions), the inclusion of exogenous explanatory variables on
volatility.




\section*{Appendix I}

This appendix presents the proof of proposition 1 of the inferential procedure in the
text.

\textbf{ Propositon 1.}

We first provide the proofs of Parts 1 to 3 relating to the basic sequential inference
of the proposed model. For ease of notation, we omit the static parameter vector \mbox{\boldmath $\varphi$} from the proofs.

\bigskip
\noindent {\bf Proof of Part 1}: \\
Assume from the hypothesis that $\lambda_{t-1}|\mbox{\boldmath $Y$}_{t-1}\sim \text{Gamma}
\left(a_{t-1}, b_{t-1}  \right)$; thus, according to \citet[Chapter 14]{bib:west1997}, $$\ln(\lambda_{t-1})|\mbox{\boldmath $Y$}_{t-1}\sim \text{Log-Gamma}\left[f_{t-1}=\gamma(a_{t-1})-\ln(b_{t-1}),q_{t-1}=\gamma^{\prime}(a_{t-1})  \right],$$ where $\gamma(b_{t-1})$ and $\gamma^{\prime}(a_{t-1})$ are the digamma and trigamma functions, respectively. Next, we approximate the log-gamma distribution
by the normal distribution in terms of the two first moments. Then, $$\ln(\lambda_{t-1})|\mbox{\boldmath $Y$}_{t-1}\dot{\sim} \text{Normal}\left(f_{t-1},q_{t-1}\right).$$

Now, we combine the above approximated distribution of $\ln(\lambda_{t-1})$ with the evolution equation $\ln(\lambda_{t})|\ln(\lambda_{t-1})\sim \text{Normal}\left(-\alpha+\phi\ln(\lambda_{t-1}),\sigma_{\eta}^2\right)$ to obtain $p(\ln(\lambda_{t})|\mbox{\boldmath $Y$}_{t-1})$. Using the properties of the multivariate normal distribution \citep{bib:harvey1989,bib:west1997}, we have
\begin{eqnarray*}
    p(\ln(\lambda_{t})|\mbox{\boldmath $Y$}_{t-1})&\dot{=}&\displaystyle\int
    p(\ln(\lambda_{t-1})|\mbox{\boldmath $Y$}_{t-1})p(\ln(\lambda_{t})|\ln(\lambda_{t-1}))d\ln(\lambda_{t-1})\\
    &\,{\buildrel d \over =}\,&\text{Normal}\left(f_{t|t-1},q_{t|t-1}\right), \text{where}
\end{eqnarray*}
$f_{t|t-1}=-\alpha+\phi f_{t-1}$ and $q_{t|t-1}=\phi^2 q_{t-1}+\sigma_{\eta}^2$.

Since $\lambda_{t}|\mbox{\boldmath $Y$}_{t-1}\sim \text{Gamma}\left(a_{t|t-1},b_{t|t-1}\right)$ and $\ln(\lambda_{t})|\mbox{\boldmath $Y$}_{t-1}\dot{\sim} \text{Normal}\left(f_{t|t-1},q_{t|t-1}\right),$ the pair $(a_{t|t-1},b_{t|t-1})$
can be elicited in terms of the two first moments
$f_{t|t-1}=\gamma(a_{t|t-1})-\ln(b_{t|t-1})$ and
$q_{t|t-1}=\gamma^{\prime}(a_{t|t-1})$. With suitable approximations for the digamma and trigamma functions \citep{bib:abramovitz1964},
we have $f_{t|t-1}\approx \ln(a_{t|t-1})-\ln(b_{t|t-1})$ and
$q_{t|t-1}\approx \frac{1}{a_{t|t-1}}$, and then $a_{t|t-1}=q_{t|t-1}^{-1}$ and
$b_{t|t-1}=\exp(-f_{t|t-1})q_{t|t-1}^{-1}$. Now, by replacing $f_{t|t-1}$ and $q_{t|t-1}$ by their respective expressions, we have $a_{t|t-1}=(\phi^2 a_{t-1}^{-1}+\sigma_{\eta}^2)^{-1}$ and $b_{t|t-1}= \frac{\exp(\alpha)(a_{t-1}/b_{t-1})^{-\phi}}{(\phi^2 a_{t-1}^{-1}+\sigma_{\eta}^2)}$.

 Therefore, $$\lambda_{t}|\mbox{\boldmath $Y$}_{t-1}\dot{\sim}\text{Gamma} (a_{t|t-1}, b_{t|t-1}), \text{where}$$
 $a_{t|t-1}=(\phi^2 a_{t-1}^{-1}+\sigma_{\eta}^2)^{-1}$ and $b_{t|t-1}= \frac{\exp(\alpha)(a_{t-1}/b_{t-1})^{-\phi}}{(\phi^2 a_{t-1}^{-1}+\sigma_{\eta}^2)}$, to complete the proof of Part 1.\\
\begin{flushright}
$\Box$
\end{flushright}

\bigskip
\noindent {\bf Proof of Part 2}:\\
To calculate the on-line or update distribution of $\lambda_t$, we have \\ $p(\lambda_{t}|\mbox{\boldmath $Y$}_{t})\propto
    p(y_{t}|\lambda_{t})p(\lambda_{t}|\mbox{\boldmath $Y$}_{t-1})\propto$$\lambda_{t}^{(a_{t|t-1}+1/r)-1}
    \exp[-\lambda_{t}(b_{t|t-1}+\psi(r)|y_{t}|^r)]$.\\

   Thus, it follows that $\lambda_{t}|\mbox{\boldmath $Y$}_{t}\sim \text{Gamma}\left(a_{t},b_{t}\right)$, where $a_{t}=a_{t|t-1}+1/r$ and
    $b_{t}=b_{t|t-1}+\psi(r)|y_{t}|^r$, completing the proof.
    \begin{flushright}
$\Box$
\end{flushright}

\bigskip
\noindent {\bf Proof of Part 3}:
 \begin{eqnarray*}
    p(y_t \vert \mbox{\boldmath $Y$}_{t-1}
    )&=&\displaystyle\int\limits_0^\infty p(y_t \vert \lambda_{t}) p(\lambda_t \vert \mbox{\boldmath $Y$}_{t-1})d\lambda_{t}\\
        &=&\frac{\left(\frac{r\Gamma(3/r)^{1/2}}{2\Gamma(1/r)^{3/2}}\right)}{\Gamma(a_{t|t-1})(b_{t|t-1})^{-a_{t|t-1}}}
    \displaystyle\int\limits_{0}^{\infty}\left[\lambda_{t}^{1/r+a_{t|t-1}-1} \exp\left(-\lambda_{t}(\psi(r)|y_{t}|^r+b_{t|t-1})\right)\right]d\lambda_{t}\\
    &=&\displaystyle\frac{\Gamma\left(1/r+a_{t|t-1}\right)\left(\frac{r\Gamma(3/r)^{1/2}}{2\Gamma(1/r)^{3/2}}\right)(b_{t|t-1})^{a_{t|t-1}}}{\Gamma(a_{t|t-1})\left(\psi(r)|y_{t}|^r+b_{t|t-1}\right)^{a_{t|t-1}+1/r}}, y_{t}\in \Re,
\end{eqnarray*}
\begin{center}
 where $a_{t|t-1}=(\phi^2 a_{t-1}^{-1}+\sigma_{\eta}^2)^{-1}$ and $b_{t|t-1}= \frac{\exp(\alpha)(a_{t-1}/b_{t-1})^{-\phi}}{(\phi^2 a_{t-1}^{-1}+\sigma_{\eta}^2)}$
\end{center}
are parameters of the prior distribution of $\lambda_t$ in Part 1 of the results.
\begin{flushright}
$\Box$
\end{flushright}

\section*{Appendix II}
\textbf{ Propositon 2.}

This appendix presents the proof of Proposition 2 of the smoothing procedure in the
text. We omit the static parameter vector \mbox{\boldmath $\varphi$} from the proof. Samples are taken
from the smoothed log-precision $\ln(\lambda)$ distribution. Consequently, we obtain samples
from the precision $\lambda$ and $h=\lambda^{-1}$ volatility distributions.

\begin{eqnarray}
\label{smoodist}
p(\ln(\lambda_{t})|\ln(\lambda_{t+1}),{\textbf Y}_{t})&=&\frac{p(\ln(\lambda_{t+1})|\ln(\lambda_{t}),{\textbf Y}_{t})\times p(\ln(\lambda_{t})|{\textbf Y}_{t})}{p(\ln(\lambda_{t+1})|{\textbf Y}_{t})} \nonumber \\
p(\ln(\lambda_{t})|\ln(\lambda_{t+1}),{\textbf Y}_{t})&\doteq& \frac{N(-\alpha+\phi\ln(\lambda_{t}),\sigma_{\eta}^2)\times N(f_t,q_t)}{N(f_{t+1|t},q_{t+1|t})} \nonumber \\
p(\ln(\lambda_{t})|\ln(\lambda_{t+1}),{\textbf Y}_{t})&\propto&  \exp[\frac{-1}{2(\frac{\phi^2}{\sigma_\eta^2}+\frac{1}{q_t})^{-1}}
 \nonumber\\
&\times&(\ln(\lambda_{t})^2-2\ln(\lambda_{t})\times(\frac{\phi^2}{\sigma_\eta^2}+\frac{1}{q_t})^{-1}\nonumber\\
&\times&(\frac{\phi(\ln(\lambda_{t+1})+\alpha)}{\sigma_\eta^2}+\frac{f_t}{q_t}))]. \nonumber\\
\nonumber
\end{eqnarray}
Therefore, $p(\ln(\lambda_{t})|\ln(\lambda_{t+1}),{\textbf Y}_{t})$ is a normal distribution, with approximate mean
$\mu_{t}^{\star}=\sigma_{t}^{2\star}
\times\left(\frac{\phi(\ln(\lambda_{t+1})+\alpha)}{\sigma_\eta^2}+\frac{f_t}{q_t}\right)$ and variance $\sigma_{t}^{2\star}=\left(\frac{\phi^2}{\sigma_\eta^2}+\frac{1}{q_t}\right)^{-1}$.
\begin{flushright}
$\Box$
\end{flushright}






\bibliographystyle{unsrtnat}
\bibliography{biblio}

\end{document}